\documentclass[aps,reprint,prr,superscriptaddress]{revtex4-1}

\usepackage[english]{babel}
\usepackage[utf8]{inputenc}      
\usepackage{csquotes}
\usepackage{todonotes}
\usepackage{comment}
\usepackage{amssymb}
\usepackage{amsmath}
\usepackage{exscale}
\usepackage{bm}
\usepackage{hyperref}
\usepackage{multirow}

\renewcommand{\d}{\text{d}}
\newcommand{\av}[1]{\left<#1\right>}
\def\bea{\begin{eqnarray}}
\def\eea{\end{eqnarray}}
\def\la{\langle}
\def\ra{\rangle}

\begin{document}


\title{Coarse-grained Second Order Response Theory}

\author{Fenna M\"uller}
\email[]{fenna.mueller@theorie.physik.uni-goettingen.de}
\affiliation{Institute for Theoretical Physics, University of G\"ottingen, Friedrich-Hund-Platz 1, D-37077 G\"ottingen, Germany}

\author{Urna Basu}
\email{urna@rri.res.in}
\affiliation{Raman Research Institute,e, Bangalore 560080, India}

\author{Peter Sollich}
\email{peter.sollich@uni-goettingen.de}
\affiliation{Institute for Theoretical Physics, University of G\"ottingen, Friedrich-Hund-Platz 1, D-37077 G\"ottingen, Germany}
\affiliation{Department of Mathematics, King's College London, Strand, London WC2R 2LS, UK}

\author{Matthias Kr\"uger}
\email{matthias.kruger@uni-goettingen.de}
\affiliation{Institute for Theoretical Physics, University of G\"ottingen, Friedrich-Hund-Platz 1, D-37077 G\"ottingen, Germany}

\date{\today}

\begin{abstract}
While linear response theory, manifested by the fluctuation dissipation theorem, can be applied at any level of coarse graining, nonlinear response theory is fundamentally of microscopic nature.
For perturbations of equilibrium systems, we develop an exact theoretical framework for analyzing the nonlinear (second order) response of coarse grained observables to  time-dependent perturbations, using a path-integral formalism. The resulting expressions involve correlations of the observable with coarse grained path weights. The time symmetric part of these weights depends on the paths and perturbation protocol in a complex manner; in addition, the absence of Markovianity prevents slicing of the coarse-grained path integral. 
We show that these difficulties can be overcome and the response function can be expressed in terms of path weights corresponding to a single-step perturbation. This formalism thus leads to an extrapolation scheme
where measuring linear responses of coarse-grained variables suffices to determine their second order response. We illustrate the validity of the formalism with an exactly solvable four-state model and the near-critical Ising model. 
\end{abstract}


\maketitle

\section{Introduction\label{intro}}
Many systems of practical and scientific relevance are of intrinsic stochastic nature with properties dominated by fluctuations, e.g.~colloidal particles, protein folding networks, molecular motors or stochastic heat engines \cite{seifert2012stochastic}. 
Such systems lend themselves to descriptions by statistical physics. A variety of approaches exist for this, including
the famous Jarzynski equation \cite{jarzynski1997nonequilibrium} and Crooks theorem \cite{crooks1999}, which concern the work done while driving a system far from equilibrium. 
In contrast, (nonlinear) response theory treats {\it arbitrary} observables, starting near equilibrium with the fluctuation-dissipation theorem, which relates the {\it linear} response to equilibrium fluctuations \cite{callen1951irreversibility,kubo2012statistical}. 
Higher orders in perturbation have also been derived, e.g., for Markov jump processes \cite{diezemann2012nonlinear}, using path integrals \cite{basufrenetic}, or  in terms of correlation functions \cite{Yamada67,Evans88,andrieux07,bouchaud2005nonlinear,lippiello08,colangeli2011meaningful,lucarini2012beyond,Kubo54}.
Nonlinear response theory has also been applied experimentally, enabling measurement of the second order response from an equilibrium average \cite{helden2017measurement}.

The above nonlinear response approaches typically rest on the assumption that all  relevant degrees of freedom (d.o.f.) are known and measurable. This is not the case in many experimental settings and one is thus faced with the additional challenge of coarse graining. 

 Taking the example of a colloidal particle in a simple solvent, the bath d.o.f.~can easily be integrated out because they relax quickly on typical colloidal timescales, and can (thus) be assumed to be in an {\it equilibrium} state \cite{seifert2008stochastic,altland}.
Approaches such as Mori-Zwanzig projection operators formalise this idea by identifying a subset of slow d.o.f.~to be relevant and integrating out the fast d.o.f. \cite{zwanzig2001nonequilibrium,te2019mori,mori1965transport}.
Indeed, fluctuation relations and response theory have been shown to hold approximately under the assumption that subsystems reach local equilibrium \cite{bravi2017statistical,rahav2007fluctuation,esposito2012}. Other types of coarse graining preserve fluctuations \cite{altaner2012} or use other physical or computational restrictions, as e.g.~in polymer physics \cite{vettorel2010fluctuating,sambriski2007theoretical} or biophysics \cite{saunders2013coarse,ayton2009hybrid}.

Adding a second colloidal particle to our example 
illustrates the next level of complexity: 
if the position of one colloidal particle is unknown, experimental estimation of potentials, entropy production and probability distributions may be incorrect, as shown experimentally in Ref.~\cite{mehl2012role}. This is the case, for example, if a driving protocol acts on the unknown degree of freedom. Such questions in relation to
entropy production, work and other thermodynamic notions in stochastic processes have been analyzed under coarse-graining, both theoretically \cite{kahlen2018hidden,rahav2007fluctuation,garcia2016thermodynamic,esposito2012}  and experimentally \cite{ribezzi2014free,mehl2012role}.
But what about the nonlinear (second order) response in coarse-grained systems? As detailed below, nonlinear orders remain challenging in coarse grained systems, even if entropy productions are found correctly. 
Ref.~\cite{basuextrapolation} developed second order response theory in a system coarse-grained to a finite number of states, proposing and verifying an extrapolation scheme for the second order response from linear contributions. 
Notably, this approach does not rely on a separation of time scales as demonstrated explicitly for a model system \cite{basuextrapolation}.
While Ref.~\cite{basuextrapolation} is restricted to  perturbations that remain constant after an initial, instantaneous jump, in this manuscript, we generalize this approach to include arbitrary time dependence.

Starting from microscopic response theory from path integrals, we derive a response theory for a finite number of coarse-grained states.
Coarse graining the path integrals yields coarse grained path weights, including entropy production, but also the more difficult time symmetric part of the corresponding weights, from which the second order response can be found. 

These formal expressions can be used in practice, e.g.~via an extrapolation scheme. In this scheme, performing a linear response experiment (or simulation) is sufficient to obtain the second order response. 
We show how to measure the second order response for any protocol from linear perturbations with one step only, thereby greatly facilitating the measurement.
This concept is illustrated and verified below in an analytically solvable jump process and in simulations of the 2d Ising model.

\section{System and Nonlinear response theory\label{theory}}
In this section we present nonlinear (second order) response theory, starting from the microscopic description, which is then coarse grained to macroscopic observables. 

\subsection{Microscopic description \label{sec:micro}}
Consider a classical system of interacting degrees of freedom, e.g. a fluid, with  phase state at time $s$ denoted by $x_s\in\Gamma$, which is in general of high dimensionality. 
Assuming that the state $x_s$ is of sufficient microscopic resolution, $(x_s)_{s \in [0,t]}$ is a Markov process.
In the absence of perturbations, the system is in equilibrium at temperature $T= 1/(k_B \beta)$, with the Boltzmann constant $k_B$.
When perturbed, the system is out of equilibrium, a situation which we aim to analyze here. 
We therefore start by reviewing an expansion of the system around equilibrium in terms of path integrals \cite{basufrenetic,wynantsthesis}.

We introduce a volume form on the space of paths $p(\omega) \mathbb{D} \omega$,
so that $p(\omega)$ is the probability (density) to find the path $\omega = \{x_s\}_{s \in [0,t]}$. 
The average of a \emph{state observable} $O(x_t)$, which depends on the state of the system at time $t$, is given by
\begin{align}
	\av{O}(t) = \int O(x_t) p(\omega) \mathbb{D} \omega~,\label{eqn:av}
\end{align}
On the r.h.s.~we have an integral over paths ending at time $t$, weighted by $p$.

Consider now a perturbation by a potential $\nu (x)$, 
acting on the system for times $s\geq 0$, 
carrying as prefactors a dimensionless perturbation strength $\varepsilon$ and  a dimensionless \emph{protocol} $h(s)$ of order unity, 
so that the full perturbation to the energy of state $x$ is given by $-\varepsilon h(s)\nu (x)$. 
The aim of response theory, as for instance developed in Refs. \cite{basufrenetic,wynantsthesis}, is to express the path probability in the perturbed non-equilibrium system, $p_{\varepsilon,h}(\omega)$, in terms of the equilibrium path probability $p_{\text{eq}}(\omega)$ and orders of the perturbation strength $\varepsilon$.
This is done in terms of a Radon-Nikodym derivative, which relates different probability measures in the Radon-Nikodym theorem \cite{wynantsthesis}.
Here, it connects the probability densities \cite{wynantsthesis} 
\begin{align}
	p_{\varepsilon,h}(\omega) = e^{-a_{\varepsilon,h}(\omega)} p_{\text{eq}}(\omega),
\end{align}
where we have introduced an \emph{action} $a$ quantifying the deviation from equilibrium.
It will be useful to consider the time reversed process, described by backward paths. These are given by $\theta \omega = \pi\{x_{t-s}\}$, where $\pi$ refers to the kinematic sign reversal, such as flipping the sign of velocities, and evolve under the reversed protocol $\bar h(s)=h(t-s)$ \footnote{We consider fields $h$ that are even under time reversal since they are scalar. If $h$ is itself odd, its sign also has to be flipped in time reversal to obtain the proper entropy production}.  
Integration over the backwards path weight, $p_{\varepsilon, \overline{h}}(\theta \omega)$, yields 
\begin{align}
  &   \int O(x_t) p_{\varepsilon, \overline{h}}(\theta \omega) \mathbb{D} \omega = 
  \int O(x_0) p_{\varepsilon,\overline{h}}(\omega) \mathbb{D} \omega \nonumber \\
  &= \av{O}(0) = \av{O}^{\text{eq}}(0) = \av{O}^{\text{eq}}(t) ~,
  \label{eqn:intout}
\end{align}
where $\av{O}^{\text{eq}}$ denotes the equilibrium average. 
Eq.~\eqref{eqn:intout} uses that the system is in equilibrium at time $s=0$: in the second term in the first line of Eq.~\eqref{eqn:intout} we can split the path integral into one over all paths starting at $x_0$, followed by an integral over $x_0$. The first integration gives the initial distribution of $x_0$, and the second one then the average of $O$ at $s=0$, which is the equilibrium average. See
appendix \ref{app:rel} for details. 
Eq.~\eqref{eqn:intout} inspires a decomposition of the action $a = d -s/2$, into its time symmetric and anti-symmetric  parts $d$ and $s$, respectivly. 
The time-anti symmetric part $s_{\varepsilon,h}=  \ln\left( p_{\varepsilon,h}(\omega)/p_{\varepsilon,\overline{h}}(\theta \omega) \right)$ is called the entropy production.
For the potential perturbation given above and assuming the time reversed process also starts in equilibrium, it has the form
\begin{align}
  \begin{split}	
  s_{\varepsilon,h}(\omega) =  \varepsilon \beta \big(& h(t) \nu(x_t) - h(0) \nu(x_0) - \int_0^t \dot{h}_s \nu(x_s) \d s \big)~,
      \end{split}
	\label{eqn:entropy}
\end{align}
as shown for specific examples in Ref. \cite{basufrenetic} and quite generally in Ref. \cite{wynantsthesis}.
The time-symmetric part $d_{\varepsilon,h}=-\frac{1}{2}\ln(p_{\varepsilon,h}(\omega) p_{\varepsilon,\overline{h}}(\theta \omega)/p^2_{\text{eq}}(\omega))$, sometimes denoted dynamical activity, depends on further details.
No explicit form can be given without specifying the dynamics of the system \cite{basufrenetic}.

Expanding in terms of the perturbation strength $\varepsilon$ and subtracting the path integral over backwards paths, as considered in Eq. \eqref{eqn:intout}, yields
\begin{align}
  \begin{split}
	&\av{O} = \av{O}^{\text{eq}} 
	+ \varepsilon \av{s_h' O}^{\text{eq}} - \varepsilon^2\av{s_h' d_h' O}^{\text{eq}} + \mathcal{O}\left( \varepsilon^3 \right)~.
      \end{split}
	\label{eqn:recap}
\end{align}
We introduced the notation $f'= \frac{\d f}{ \d \varepsilon}|_{\varepsilon=0}$ so that $s'$ is immediately found from Eq.~\eqref{eqn:entropy}.
The derivative $d'$ of the time-symmetric component  is given in terms of the derivative of $p$, \cite{wynantsthesis},
\begin{align}
        d_{h}'(\omega) = -\frac{1}{2p_{\text{eq}}(\omega)} \left( p_{h}'(\omega) + p_{\overline{h}}'(\theta \omega) \right).
        \label{eqn:dp}
\end{align}
Examples for different dynamics may be found in Refs.~\cite{basufrenetic,wynantsthesis}.

We finally introduce a notation for the $n$-th order  response of the non equilibrium average $\av{O}$, \footnote{This differs from previous notations in Ref. \cite{helden2017measurement,basufrenetic}, where these quantities are denoted by $\chi^{(n)}$ or $\chi_n$.}
\begin{equation}
  \av{O}^{(n)} := \frac{1}{n!}\frac{\d^n}{\d \varepsilon^n} \av{O}|_{\varepsilon=0},
\end{equation}
which we analyze up to $n=2$ in this manuscript.

It is important to note that Eq.~\eqref{eqn:recap} relies on the unperturbed system being in equilibrium, via detailed balance. For non-equilibrium cases, already the linear response involves $d'_h$ \cite{baiesi2009fluctuations}, and the second order involves $d''_h$. This would lead to differences in the coarse grained relations derived below. 
\subsection{Coarse-grained description \label{sec:coarse}}
We now turn to a coarse-grained version of the stochastic process introduced above. 
This is motivated by the fact that experimental resolution is always limited, so that in general only coarse grained observables can be monitored. 
Furthermore, developing non-equilibrium thermodynamics for macroscopic variables is an important goal of statistical physics. The coarse graining as performed here allows for a practical extrapolation scheme, as detailed below.

We thus consider a countable number of coarse-grained, discrete (stochastic) states $X_s\in\Gamma'$, 
with a function $\varphi$ uniquely mapping $\Gamma$ to $\Gamma'$, i.e. $X_s = \varphi(x_s)$.  
\begin{figure}[h]
	\centering
	\includegraphics[width=0.45\textwidth]{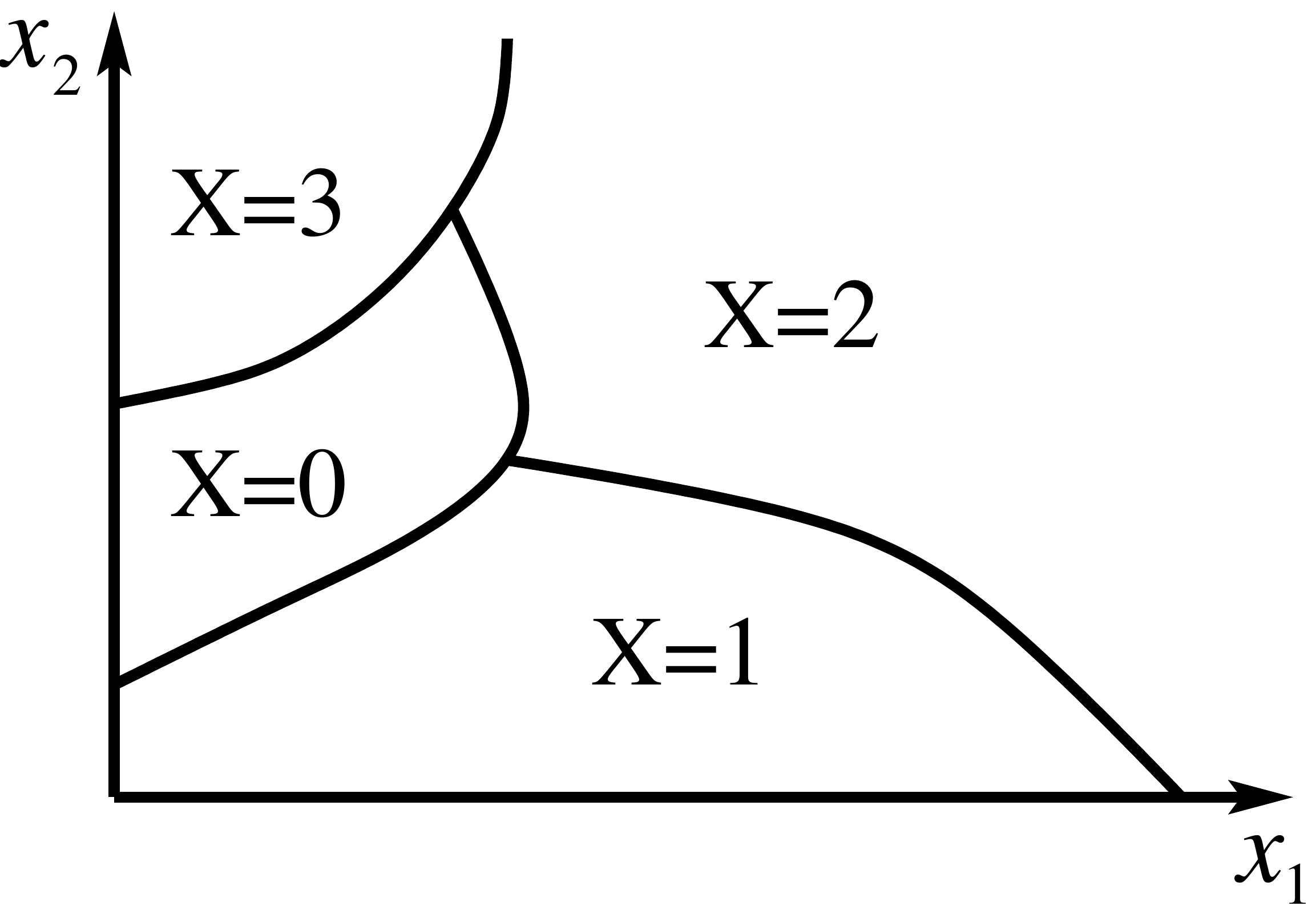}
	\caption{
	  Illustration of coarse graining: A continuous microscopic phase space $\Gamma$ (here spanned by $x_1$ and $x_2$) is coarse grained into states $X=0,1,2,3$ that make up the coarse grained space $\Gamma'$.}
	\label{fig:example}
\end{figure}

Figure \ref{fig:example} illustrates this mapping of a microscopic continuous state space $\Gamma$ to $\Gamma' = \{0,1,2,3\} $ consisting of four coarse-grained states.
Note that this approach does not rely on a separation of time scales of slow and fast variables and thus remains valid even when the coarse-grained process is not  Markovian.
The spirit of this coarse graining is hence distinct from the idea that underlies typical approximations based on  local equilibrium. 

A crucial {\it physical} assumption or requirement is that the  perturbation  potential $\nu$ acts on the  coarse level as well, in other words, $\nu$ is a function of coarse-grained states. 
In the introductory example of  colloids in a solvent, it means that the perturbation acts only on those colloids whose positions are being monitored. 
To emphasize this, we introduce a potential $V(X)$ acting on coarse-grained states, 
so that $\nu(x) = V(X) $ for all $x$ satisfying $\varphi(x) = X$.
Importantly, under this assumption, the  entropy production is a functional of coarse-grained paths $\Omega = \{X_s\}_{s \in [0,t]}$ (see Ref.~\cite{rahav2007fluctuation} for a statement in a similar spirit). 
We define the coarse-grained entropy production;
\begin{align}
	\begin{split}
	  S_{h}'(\Omega) &:= \frac{1}{P_{\text{eq}}(\Omega)}\int_\Omega s_h'(\omega) p_{\text{eq}}(\omega) \mathbb{D} \omega \\
	&=  \beta \big( h(t) V(X_t) - h(0) V(X_0) - \int_0^t \dot{h}(s) V(X_s) \d s \big)~.
\end{split}
	\label{eqn:macroent}
\end{align}
The index $\int_\Omega$ indicates that the integral runs over all micro paths belonging to the coarse-grained path $\Omega$, and  
$P_{\text{eq}}(\Omega)$ is the equilibrium weight for path $\Omega$. Eq.~\eqref{eqn:macroent} follows directly from Eq.~\eqref{eqn:entropy}, by noting that $\nu(x) = V(X) $ for  $x$ with $\varphi(x) = X$. 
The {\it linear} response of a state observable $O(X_t)$ is thus
\begin{align}
        &\av{O}^{(1)} =  \av{S_h' O}^{\text{eq}}.
        \label{eqn:lin}
\end{align} 
Comparing with Eq.~\eqref{eqn:recap}, we see that the formalism of linear response is the same when applied to a microscopic or any coarse-grained description, so that the process of coarse graining is easily performed for the linear response.

In order to obtain the second order response, we coarse grain the second order response Eq.~(\ref{eqn:recap}), making use of the coarse-grained entropy production given in Eq.~\eqref{eqn:macroent},

\begin{align}
	\av{O}^{(2)}[h](t) &=  -\int s_{h}'(\omega) d_{h}'(\omega) O(X_t) p_{\text{eq}}(\omega)\mathbb{D} \omega \nonumber\\
	&=  - \int S_{h}'(\Omega) \left[\int_\Omega  d_{h}'(\omega) p_{\text{eq}}(\omega)\mathbb{D} \omega \right]O(X_t) \mathbb{D} \Omega \nonumber\\
	&= - \int S_{h}'(\Omega)  D_h'(\Omega) O(X_t) P_{\text{eq}}(\Omega) \mathbb{D} \Omega~.
	\label{eqn:chi2coarse}
	\intertext{Here we have identified the coarse-grained $D'$ as the average of $d'$ over micropaths belonging to $\Omega$}
	& D_h'(\Omega)P_{\text{eq}}(\Omega) := \int_\Omega  d_{h}'(\omega) p_{\text{eq}}(\omega)\mathbb{D} \omega,\label{eqn:macrod}
	 \end{align}
 similarly to the coarse-grained entropy production of Eq.~\eqref{eqn:macroent}.
In Eq.~\eqref{eqn:chi2coarse} we split the integration over microscopic paths by first integrating over paths belonging to a given coarse grained path $\Omega$ and subsequently integrating over the latter. We then used that the  entropy production takes the same value for all microscopic paths $\omega$ belonging to the same $\Omega$, so that $s'_h$ can be taken out of the integral over $\mathbb{D} \omega$ in the second line of Eq.~\eqref{eqn:chi2coarse}, thereby turning into $S'_h$.
This is why the macroscopic parts $S'_h(\Omega)$ and $D'_h(\Omega)$ of the action factorize, with the consequence that Eq.~\eqref{eqn:chi2coarse} takes a form similar to  Eq.~(\ref{eqn:recap}).
The coarse-grained path integral appearing here can be written as
\begin{align}
  \int \mathbb{D} \Omega = \sum_{X_{t_1}\in\Gamma'} \dots \sum_{X_{t_N}\in\Gamma'}~,\label{eqn:cgint}
\end{align}
by discretizing time into $N$ lattice points and making use of the discrete nature of $X$. 
Other ways of representing $\int \mathbb{D} \Omega$ can be found in Appendix \ref{app:pi}.
$D'_h(\Omega)$ in Eq.~\eqref{eqn:macrod} can be written, using \eqref{eqn:dp}, as
\begin{align}
  \begin{split}
        &D_{h}'(\Omega) P_{\text{eq}}(\Omega) = \int_\Omega -\frac{1}{2} \left( p_{h}'(\omega) + p_{\overline{h}}'(\theta \omega) \right) \mathbb{D} \omega\\
        &= -\frac{1}{2} \left( P_{h}'(\Omega) + P_{\overline{h}}'(\theta \Omega) \right),
      \end{split}
        \label{eqn:Dp}
\end{align}
where we have introduced (derivatives of the) non-equilibrium weight $P_{h}'(\Omega)$.
One important difference between Eqs.~\eqref{eqn:Dp} and \eqref{eqn:dp} lies in the Markov property of $x$, which is absent for $X$: 
while $p_{\varepsilon,h}(\omega)$ can be cut into (temporal) pieces according to the Chapman-Kolmogorov-Equation, this is not possible for $P_{\varepsilon,h}(\Omega)$.

The main challenge that remains is the determination of $D'(\Omega)$. How $D'(\Omega)$ appears in practice will be analyzed in Section \ref{sec:conc} by decomposing the time dependence of the protocol into a number of discrete steps.  
Section~\ref{sec:fourstate} verifies and illustrates these findings via analytical solutions of a four state model, and section~\ref{sec:recipe} will apply the extrapolation scheme to the Ising model.

\section{From stepwise perturbation to the second order susceptibility}\label{sec:conc}
Eq. (\ref{eqn:chi2coarse}) describes the second order response $\av{O}^{(2)}$ in terms of a path integral with linear contributions at most.
However, evaluating the path integral holds the challenge of finding $D'(\Omega)$ in Eqs.~\eqref{eqn:macrod} or \eqref{eqn:Dp}. 
In this section we demonstrate that, starting with protocols of finite number of discrete steps,  $D'(\Omega)$ turns into a tensor of finite order.
We discuss the simplifications arising if the coarse grained process is itself Markovian in Appendix \ref{sec:markov}.
\subsection{A single step perturbation}\label{sec:si}
The case of a perturbation with a single step in time, i.e., 
\begin{align}\label{eqn:hs}
h(s)= \Theta(s),
\end{align}was considered in Ref.~\cite{basuextrapolation}, and for the sake of completeness we summarize the derivation here. The entropy production reads in this case
\begin{align}\label{eqn:Ssingle}
        S'(\Omega)= \beta(V(j)-V(i)) = S_{ij}'.
\end{align}
We have introduced the abbreviations $i=X_0$ and $j=X_t$ for the states at times $0$ and $t$. 
The above form of $S'$ reduces the path integral in Eq.~\eqref{eqn:chi2coarse} into a sum of terms \cite{basuextrapolation}: \begin{align}
        \begin{split}
        &\av{O}^{(2)}(t)
	= - \int S'(\Omega)  D'(\Omega) O(X_t) P_{\text{eq}}(\Omega) \mathbb{D} \Omega\\
&= - \sum_{ij\in \Gamma'}
        S'_{ij}\int_{ij} 
        D'(\Omega) O(X_t) P_{\text{eq}}(\Omega) \mathbb{D} \Omega\\
        &= -\sum_{ij} S_{ij}' D_{ij}' P_{ij}^{\text{eq}} O(j)~.
      \end{split}
        \label{eqn:onestep}
      \end{align}
Here, $\int_{ij}$ is a path integral with fixed start and end states $i$ and $j$, 
which yields the joint probability
\begin{align}
P_{ij}=\int_{ij}   P(\Omega) \mathbb{D} \Omega~ .
\end{align}
According to Eq.~\eqref{eqn:Dp}, the time-symmetric component is given by
\begin{align}
D_{ij}' P_{ij}^{\text{eq}} &= 
        \int_{ij} 
D'(\Omega)  P_{\text{eq}}(\Omega) \mathbb{D} \Omega\\
        &= -\frac{1}{2} 
        \left( P_{ij}' + P_{ji}' \right)
\end{align}
For the step perturbation of Eq.~\eqref{eqn:hs}, the protocol equals its reverse and the protocol reversal appearing in Eq.~\eqref{eqn:Dp} is obsolete.
We note that, for a single step, $D'(\Omega)$ turns into a matrix $D_{ij}'$, which is related to the linear response of the coarse-grained probability $P_{ij}$. 
The latter can be measured easily, giving rise to the extrapolation scheme introduced in Ref.~\cite{basuextrapolation}, which we discuss further in section \ref{sec:recipe} below. 

\subsection{A two step perturbation \label{sec:twostep}} 

We add one more step to the protocol at time $0\leq \tau \leq t$,  introducing the corresponding state $k=X_{\tau}$.
Denoting the step sizes by $\Delta h_0$ and $\Delta h_1$, the protocol is then 
\begin{align}
h(s) = \Delta h_0 \Theta(s) + \Delta h_1 \Theta(s-\tau). \label{eq:h_2step}
\end{align}
Recalling the definition of $S_{ij}'$ and $S'(\Omega)$ in Eq.~\eqref{eqn:Ssingle}, Eq.~\eqref{eqn:macroent} yields the entropy production for two steps 
\begin{align}
        S'(\Omega)=  & \Delta h_1 S_{kj}' + \Delta h_0 S_{ij}'  =: S_{ikj}'.
\label{s_ikj}
\end{align}
Similarly to Eq.~\eqref{eqn:onestep}, the path integral turns into sums over states at times $0, \tau$ and $t$,
\begin{align}
        \begin{split}
        &\av{O}^{(2)}(t)
        = - \int S'(\Omega)  D'(\Omega) O(X_t) P_{\text{eq}}(\Omega) \mathbb{D} \Omega\\
	&= - \sum_{i,k,j \in \Gamma'} S_{ikj}' 
        \int_{ikj} 
	D'(\Omega) O(X_t) P_{\text{eq}}(\Omega) \mathbb{D} \Omega\\
	&= -\sum_{ikj} S_{ikj}' D_{ikj}' P_{ikj}^{\text{eq}} O(j)~.      
\end{split}
        \label{eqn:twosteps}
      \end{align}
      Consistent with the notation above, the path integral $\int_{ikj} \mathbb{D} \Omega$ is restricted to paths passing through the states $i,k,j$ and
      yields the probability of being in state $i$ at time $s=0$, in $k$ at time $\tau$ and in state $j$ at the time of the measurement $t$:
      \begin{align}
	P_{ikj} = \int_{ikj} P(\Omega) \mathbb{D} \Omega~.
      \end{align}
 Applying this notation, we can identify the time-symmetric contribution by integrating in Eq.~\eqref{eqn:Dp}
\begin{align}
&D_{ikj}' P_{ikj}^{\text{eq}}
              = -\frac{1}{2}
        \left( P_{ikj}' + \overline{P_{ikj}}' \right)~.
\end{align}
where we introduced the probability $\overline{P_{ikj}}$ under time and protocol reversal.
More specifically $\overline{P_{ikj}}$ is the probability to measure $j$ at time $s=0$, $k$ at time $s=t-\tau$ and $i$ at time $s=t$ under the backwards protocol $\overline{h}(s) = (\Delta h_0 + \Delta h_1)\Theta(s)-\Delta h_1 \Theta(s-(t-\tau)) $.
In order to arrive at $\overline{P}$ we have swapped the two operations of time reversal and integration over coarse-grained paths.

By construction, $D'$ must be linear in the protocol $h$, so that it can be decomposed into
\begin{align}
&D_{ikj}'=
\Delta h_0 D_{ikj}'[\Theta_0] + \Delta h_1 D_{ikj}'[\Theta_{\tau}].
      \label{eqn:D}
      \end{align} 
The quantities $D_{ikj}'[\Theta_s]$ appearing on the right give the value of $D_{ikj}'$ for a {\em single} perturbation step at time $s$, e.g.~$D_{ikj}'[\Theta_0]$ is extracted for a perturbation consisting of a step at time $s=0$.
We have thus obtained a tensor $D'$ with three indices, which as before is connected to coarse-grained probabilities $P$
\footnote{An expression of the two forms appearing in Eq.~\eqref{eqn:D} in terms of one step probabilities may be found in Appendix~\ref{sec:ApA} in Eq.~\eqref{eqn:Dt}.
}.

\subsection{Second order susceptibility for any protocol}
In this section, we consider the response to a general protocol $h$ by deriving a formula for the second order susceptibility in terms of ``one step probabilities''.
Given that the unperturbed equilibrium systems is invariant under time translations, we can express the second order response for a protocol $h$ in terms of the second order susceptibility $\chi$, \cite{findley}
\begin{align}
	\av{O}^{(2)}(t) &= \int_0^t \int_0^t \dot{h}(t_1) \dot{h}(t_2) \chi(t-t_1,t-t_2) \d t_1 \d t_2.
	\label{eqn:assumption}
\end{align}
The $\chi(t,t-\tau)$ defined in this way can be determined from the second order response to a protocol with one and two steps, since the time derivatives of such a protocol are $\delta$ distributions at the times where the field jumps.
We may thus find $\chi$ from the relations given in the previous subsections.
Comparing the definition of the second order susceptibility, Eq.~(\ref{eqn:assumption}), to the response formula given in terms of indices (\ref{eqn:twosteps}), and using the linearity of $D'$ in Eq.~\eqref{eqn:D} yields 
\begin{align}
	\begin{split}
	\chi(t,t-\tau) =  -\frac{1}{2} \sum_{ikj} &\big( S_{kj}' D_{ikj}'[\Theta_0]\\
	&+ S_{ij}'  D_{ikj}'[\Theta_{\tau}] \big)P_{ikj}^{\rm eq} O(j),
	\end{split}
	\label{eqn:1}
\end{align}
which is valid for $0 \leq\tau \leq t$. 
The case of arbitrary times $\chi(t_1,t_2)$ is obtained by inserting the respective arguments $\tau = t_1-t_2$ and $t=t_1$.

Eq.~\eqref{eqn:1} is an intermediate result: it gives the second order response function $\chi$ for any arguments in terms of sums over indices of the tensors $S_{ij}'$ and $D_{ikj}'$ obtained in Eqs.~\eqref{eqn:D0}, \eqref{eqn:Dt} and \eqref{eqn:Ssingle}. Note that as~\eqref{eqn:1} relates to different times, it comes from the cross-terms in the product of~\eqref{s_ikj} and~\eqref{eqn:D}. This is why e.g.\ the first term contains a factor $S'_{kj}$ relating to the state $k$ at time $\tau$, combined with $D'_{ikj}[\Theta_0]$ for a perturbation at time zero.

As a final simplification we note that the entropy production in Eq.~\eqref{eqn:1} carries only two indices so that we can sum over the remaining index. This sum eliminates one index from the expressions $D_{ikj}'P_{ikj}^{\text{eq}}$, which always occur jointly (compare Eqs.~\eqref{eqn:D0} and \eqref{eqn:Dt}).
One term is given by summing over the center index $k$ in the probabilities $\sum_k P_{ikj}'(\tau,t) = P_{ij}'(t)$. 
We add notation to include time and protocol, as these are varied below. 
For example $P_{ikj}[\Theta_{s_1}](\tau,t)$ denotes the probability of measuring state $i$ at time $s=0$, state $k$ at time $\tau$ and state $j$ at time $t$ under a perturbation switched on at time $s_1$.
With this notation in mind, the summation over $k$ yields
\begin{align}
  &\sum_k D_{ikj}'[\Theta_{\tau}](\tau,t) P_{ikj}^{\text{eq}}(\tau,t)\notag\\
&=  -\frac{1}{2} \left[P_{ij}'[\Theta_{\tau}](t)+P_{ji}'[\Theta_0](t)
-P_{ji}'[\Theta_{t-\tau}](t)\right]\notag\\
&=D'_{ij}[\Theta_{\tau}](t)P_{ij}^{\text{eq}}(t),\label{eqn:sumup0}
\end{align}
where we used the linearity of $P_{ji}'[\Theta_0 -\Theta_{t-\tau}](t)= \overline{P_{ij}'[\Theta_\tau](t)}$ in the backwards protocol, as given in Eq.~\eqref{eqn:Dt}.
The second term involves summation over the first index,
\begin{align}
  &\sum_{i} D_{ikj}'[\Theta_0](\tau,t) P_{ikj}^{\text{eq}}(\tau,t)\nonumber\\
&= -\frac{1}{2} \left( P_{kj}'[\Theta_{-\tau}](t-\tau) + P_{jk}'[\Theta_0](t-\tau) \right)\notag\\ 
 &=D'_{kj}[\Theta_{-\tau}](t-\tau)P_{kj}^{\text{eq}}(t-\tau).
\label{eqn:sumup1}
\end{align}
Notably, the perturbation in Eq.~\eqref{eqn:sumup1} starts at negative times $-\tau<0$ and probabilities cover a time interval of $t-\tau$ between measurements due to integrating out the first state.  
After renaming indices, we finally obtain for the second order susceptibility,
\begin{align}
  \begin{split}
 & \chi(t,t-\tau) =  -\frac{1}{2} \sum_{ij}S_{ij}'  O(j)\times  \\
  & \left(D_{ij}'[\Theta_{-\tau}](t-\tau) P_{ij}^{\text{eq}}(t-\tau)+D_{ij}'[\Theta_{\tau}](t) P_{ij}^{\text{eq}}(t)\right)~.
\end{split}
\label{eqn:twoindex}
\end{align}
This expression is symmetrical under exchange of its arguments $t_1 = t$ and $t_2 = t-\tau$, as expected from the definition of the susceptibility, Eq. \eqref{eqn:assumption}. 
For the single step perturbation the second order is given by the response function for equal time arguments $\av{O}^{(2)}[\Theta_0] = \chi(t,t)$, i.e.~by setting $\tau = 0$. The two terms in Eq.~\eqref{eqn:twosteps} then become the same and we recover Eq.~\eqref{eqn:onestep} as it should be.

 \begin{figure}[t]
	\includegraphics[width=0.4\textwidth]{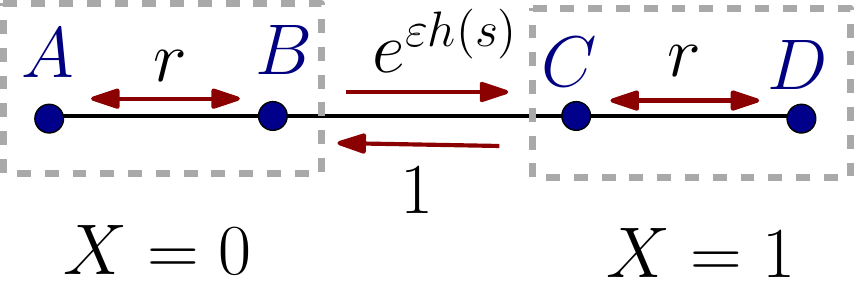}
\caption{\label{fig:4st}Illustration of coarse graining the Markov four state model into a non-Markovian two state model. Arrows denote the transitions with the given rates.}
 \end{figure}

\section{Illustration and verification: The four state model\label{sec:fourstate}}

In this Section we use a simple example system, namely, a (driven) four state model that can be solved analytically, to verify and illustrate the concepts introduced in Section \ref{sec:conc}.

\subsection{Model and coarse graining} 
The second order response can be expressed in terms of the  entropy production and a time-symmetric component, as in Eqs.~(\ref{eqn:chi2coarse}), (\ref{eqn:1}). As a proof of concept we consider 
 a Markov jump process with four states $\Gamma = \{A,B,C,D\}$, see Fig.~\ref{fig:4st}, which is then coarse grained to a two state one. Such Markov jump processes may be used to describe a variety of systems, see e.g.~Ref. \cite{wynantsthesis}.

 The conditional probabilities $p_{\alpha\delta}(s,t)$ for occupying state $\delta$ at time $t$ if occupying $\alpha$ at an earlier time $s$ is described by the Master equation (an equivalent equation holds for occupation densities) 
\begin{align}
	\frac{\partial }{\partial t} p_{\alpha \delta}(s,t) = \sum_{\gamma \in \Gamma} p_{\alpha \gamma}(s,t) q_{\gamma \delta}(t) \label{eqn:master}
\end{align}
with the rate $q_{\alpha \delta}(t)$ for the transition from the state $\alpha$ to $\delta$ and setting $q_{\alpha \alpha} = - \sum_{\delta \not= \alpha} q_{\alpha \delta}$ to ensure  probability conservation, c.f.~Ref.~\cite{wynantsthesis}. 

More explicitly, we use time-independent rates $q_{AB}= q_{BA} = q_{CD} = q_{DC} =r$ for the side links, and the center links have rates $q_{BC}(t) = e^{\varepsilon h(t)}$ and $q_{CB} = 1$ (with all other rates being 0).
Only the rate of the center link $q_{BC}(t)$ is time dependent and this will be used to drive the system.
The rate matrix $q(s)$ is given explicitly in the appendix, Eq.~\eqref{eqn:rates}. We use $\beta=1$ in this section.

\begin{figure}[t]
  \centering
  \includegraphics[width=0.45\textwidth]{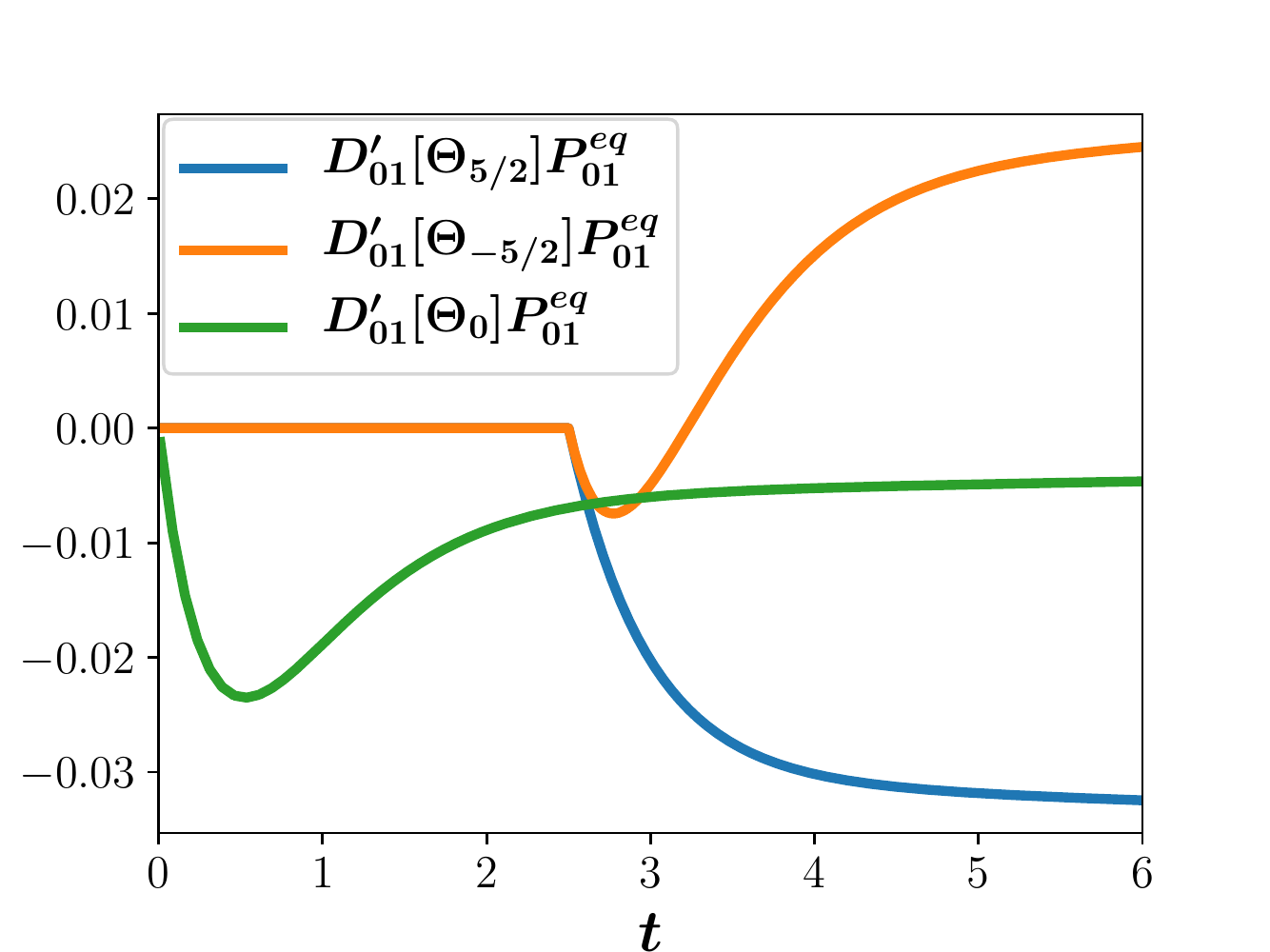}
  \caption{The time symmetric components $D_{01}'$  contributing to the second order susceptibility $\chi(t,t-\tau)$ and $\chi(t,t)$ in the four state model, Eq.~\eqref{eqn:k} with rate $r=0.1$ and $\tau =5/2$. Time arguments are chosen as they appear in Eq.~\eqref{eqn:k}.
  Due to causality, the time symmetric component vanishes for negative time arguments, $D_{01}'[\Theta_{\tau}](t) = D_{01}'[\Theta_{-\tau}](t-\tau)= 0$ for $t<\tau$. }
  \label{fig:D01}
\end{figure}

\begin{figure}[t]
	\centering
	\includegraphics[width=0.5\textwidth]{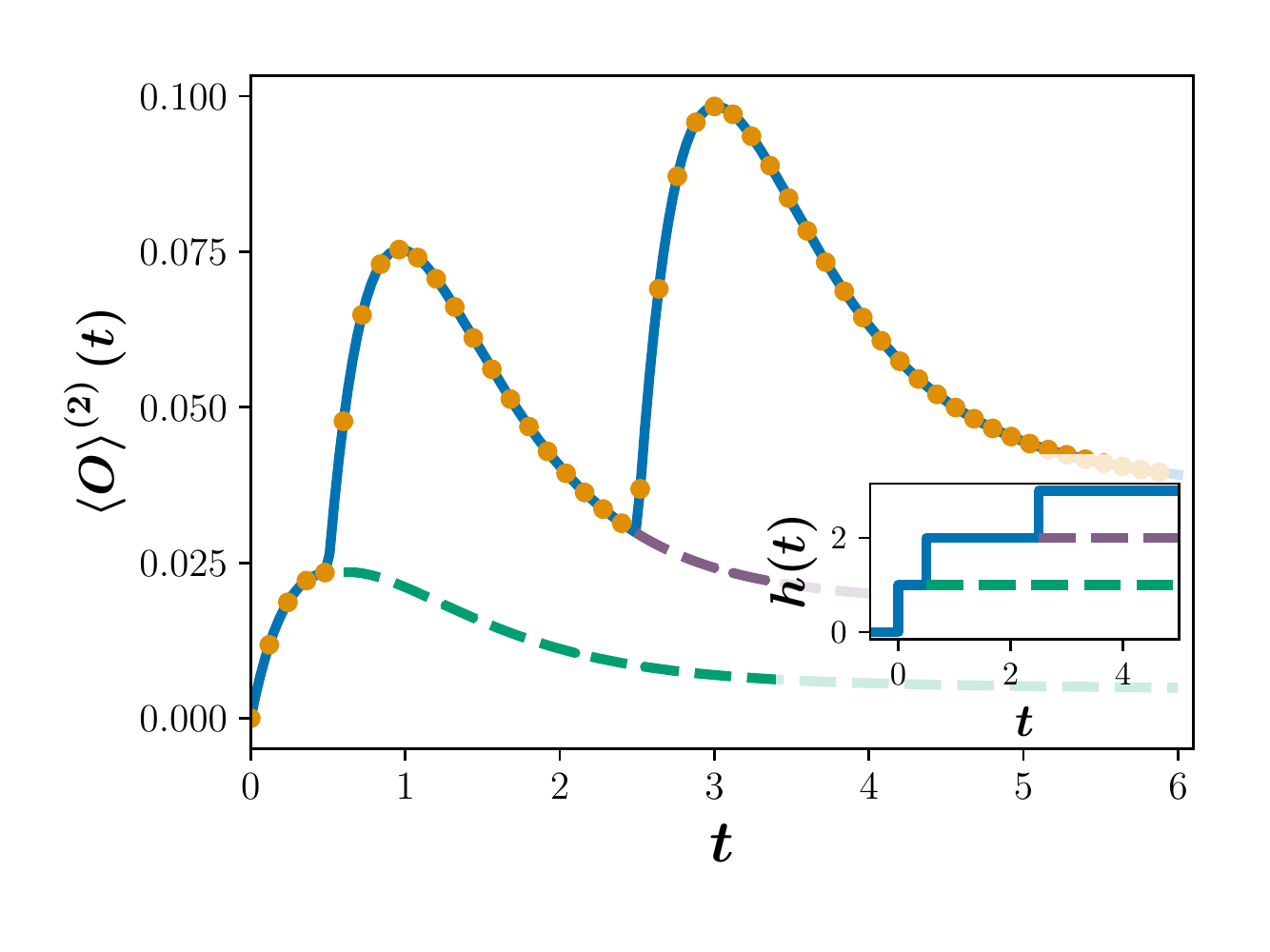}
	\caption{
	The second order response $\av{O}^{(2)}$ of the coarse grained four state model, for a driving protocol containing one, two, and three instantaneous steps (protocols shown in the inset), as a function of time $t$. Lines show the analytical solution (see Appendix \ref{app:4st}) and data points are found from Eq.~(\ref{eqn:1}) as as well analytically. 
	The `internal' rate is set to $r =0.1$, which corresponds to a strongly non-Markovian case, as also reflected in the curves: The response shows two distinct time scales of relaxation. 
    States, rates and time are naturally dimensionless. As $t\to\infty$, the curves approach zero, as, for symmetry reasons, the system has no static second order response.
}
	 	\label{fig:3steps}
\end{figure}

\begin{figure}[t]
	\centering
	\includegraphics[width=0.5\textwidth]{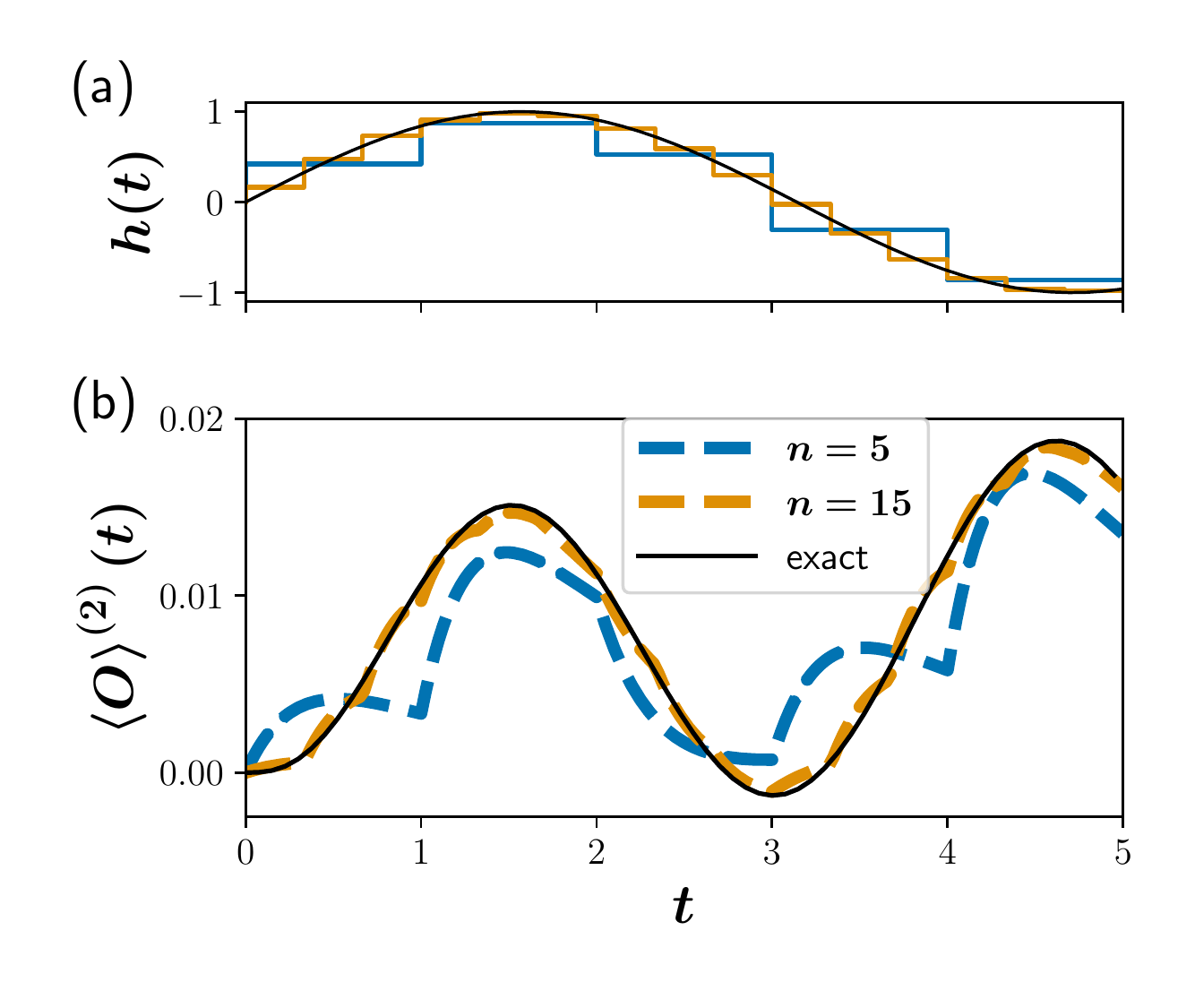}
	\caption{
	  The second order response $\av{O}^{(2)}$ of the coarse grained four state model as a function of time $t$, for a sinusoidal protocol of different temporal resolution, as shown in the upper part:
	The  sine (black) is approximated by step functions with $n$ steps of height $\Delta h_{k} = \left(\sin((k+1) \delta t) - \sin\left( (k-1) \delta t \right)  \right)/2$ and
	time increments $\delta t = t_{\rm max}/n$, here for $t_{\rm max} = 5$ and for $n =5$ (blue) and $n=15$ steps (orange).
	The lower panel shows the corresponding second order response 
	in the same colors as the corresponding protocols.  
Rates and time are given in dimensionless units.}
\label{fig:sine}
\end{figure}

This system is {\it coarse-grained} into two states $X =0$ and $X=1$ by assigning $\varphi(A) = \varphi(B) =0$ and $\varphi(C) = \varphi(D) = 1$.
The two coarse-grained states are connected by the center link $BC$ and the associated rates $q_{BC}$ and $q_{CB}$ of the underlying Markov process, which yields a non-Markovian two state process. 
Notably, in the limit $r\gg1$ the resulting two state process is Markovian, while it is strongly non-Markovian in the opposite limit $r\ll1$. 
Choosing $r=0.1$ as in Figures \ref{fig:3steps} and \ref{fig:sine} results in the system being in the latter regime.

The associated 
potential is $\nu(A)=\nu(B)=0$ and $\nu(C)=\nu(D)=1$, hence  $V(0) = 0$ and $V(1) = 1$. The rates thus fulfill $\frac{q_{BC}(t)}{q_{CB}} = e^{ \varepsilon h(t) \left( \nu(C) - \nu(B) \right)}$, which is called the microscopic reversibility condition \cite{crooks1999} or local detailed balance \cite{wynantsthesis}. This is a sufficient condition to have an  entropy production of the form given by Eq. (\ref{eqn:entropy}) \cite{wynantsthesis}.
For single step perturbations the Master equation can be solved analytically.
For more complex protocols, the solution is formally given by a time-ordered exponential, which may be expanded in orders of $\varepsilon$ using a Dyson-expansion. 
This allows us to illustrate our approach of computing the second order susceptibility from linear quantities analytically. 

\subsection{One, two, and three steps}
We compute the second order susceptibility $\chi(t,t-\tau)$ from linear contributions $S_{ij}'$ and $D_{ij}'P_{ij}^{\text{eq}}$ for perturbations switched on at different times $\pm \tau$, according to Eq.~\eqref{eqn:twoindex}.
Eq. \eqref{eqn:twoindex}  for the average in the coarse-grained two state system with $O(j) = j$ reduces to
\begin{align}
  \begin{split}
  \chi(t,t-\tau) = -\frac{1}{2} \big(& D_{01}'[\Theta_{-\tau}](t-\tau) P_{01}^{\text{eq}}(t-\tau) \\
  &+ D_{01}'[\Theta_\tau](t) P_{01}^{\text{eq}}(t)\big)~.
\end{split}
  \label{eqn:k}
\end{align}
The entropy production that contributes is $S_{01}'=1$ and the relevant time-symmetric components $D_{01}'$ for different perturbations are shown in Fig.~\ref{fig:D01}.
The explicit form of the second order susceptibility in the four state model is given in the appendix, Eq.~(\ref{eqn:kfour}).
Employing this function together with Eq.~(\ref{eqn:assumption}) enables prediction of the second order response for arbitrary protocols $\av{O}^{(2)}$.

Here, we demonstrate this by means of a protocol $h = \Theta_0 + \Theta_{1/2} + \Theta_{5/2}$ with three steps.  
As shown in Figure \ref{fig:3steps}, the response formula for the second order $\av{O}^{(2)}$ coincides with the explicit solution.
This is an example of employing the second order susceptibility from linear contributions to correctly predict the second order response under a protocol with several steps.
\subsection{Continuous protocol: Exact and discretized}
As noted above, Eq. (\ref{eqn:assumption}) readily describes any protocol, which we further illustrate by using a protocol of a sinusoidal oscillation of the form
\begin{align}
h(s)=\sin(s) \Theta_0(s).\label{eqn:sin}
\end{align}
The resulting response is shown in Fig.~\ref{fig:sine}. 
In addition to the response corresponding to the protocol of Eq.~\eqref{eqn:sin}, we show the response to the discretized versions of the protocol in the upper part of Fig.~\ref{fig:sine}. 
This illustrates the possibility of an additional coarse graining along the time axis of a protocol, which is one natural way of implementing Eq.~\eqref{eqn:twoindex} in practice (see also Sec.~\ref{sec:recipe} below). 
How fine a discretization is needed? As seen in the graph, discretizing with  an increment of unity (resulting in $n=5$ steps in the given time range) yields pronounced deviations from the exact result. On the other hand, discretizing with an increment of $1/3$, resulting in $n=15$ steps, produces a more precise approximation. This can be understood from the curves in Fig.~\ref{fig:3steps}, where the (shortest) relaxation time, or the response time, is of the order of unity. This analysis suggests that the time increment should be small compared to that response time to accurately resolve the perturbation protocol.  

\section{Extrapolation: Ising model \label{sec:recipe}}

In this section we illustrate the validity of the time-dependent coarse-grained response theory for an interacting system with many degrees of freedom, 
using the example of a near-critical Ising model.
Let us consider a  2-d lattice of size $L \times L$ with periodic boundaries; each lattice site $i$ contains a spin $\eta_i = \pm 1$ which interacts with its nearest-neighbor spins. Let the coarse grained variable $X$ correspond to a single site, say site $k$, so that we have $X=\frac 12(1+\eta_k)$. In other words, all spins except for spin $k$ will be coarse grained away, and play the role of a complex (non-Markovian) bath. This scenario may mimick the experimental situation where a system is perturbed and monitored at a local position in space. We thus introduce a magnetic field, which acts on the spin $k$, i.e.~a potential $V(X)=\eta_k=2X-1$. The Hamiltonian describing the system at any time $s$ is (setting the spin coupling to unity)
\begin{equation}
{\cal H} = -\sum_{\langle ij \rangle }{\eta_i \eta_j} - (g+ \varepsilon h(s)) \eta_k.
\end{equation}
The explicit time dependence assigned to the magnetic field via the protocol $h(s)$ gives rise to a perturbation of the system from its equilibrium state.

In the absence of the magnetic field, i.e.~with $g=h(s)=0,$ the system shows a paramagnetic to ferromagnetic transition at temperature $T_c=2.269$ in the limit of thermodynamically large size $L$ (having set the Boltzmann constant to unity). Here we consider a system of size $L=16$ at a slightly super-critical temperature $T=2.45.$ This finite sized system shows a non-zero magnetization at this temperature, which randomly flips its sign on a slow time-scale. We thus expect the resulting bath for the tagged spin $\eta_k$ to be highly non-Markovian.

\begin{figure}[t]
	\centering
	\includegraphics[width=7.5 cm]{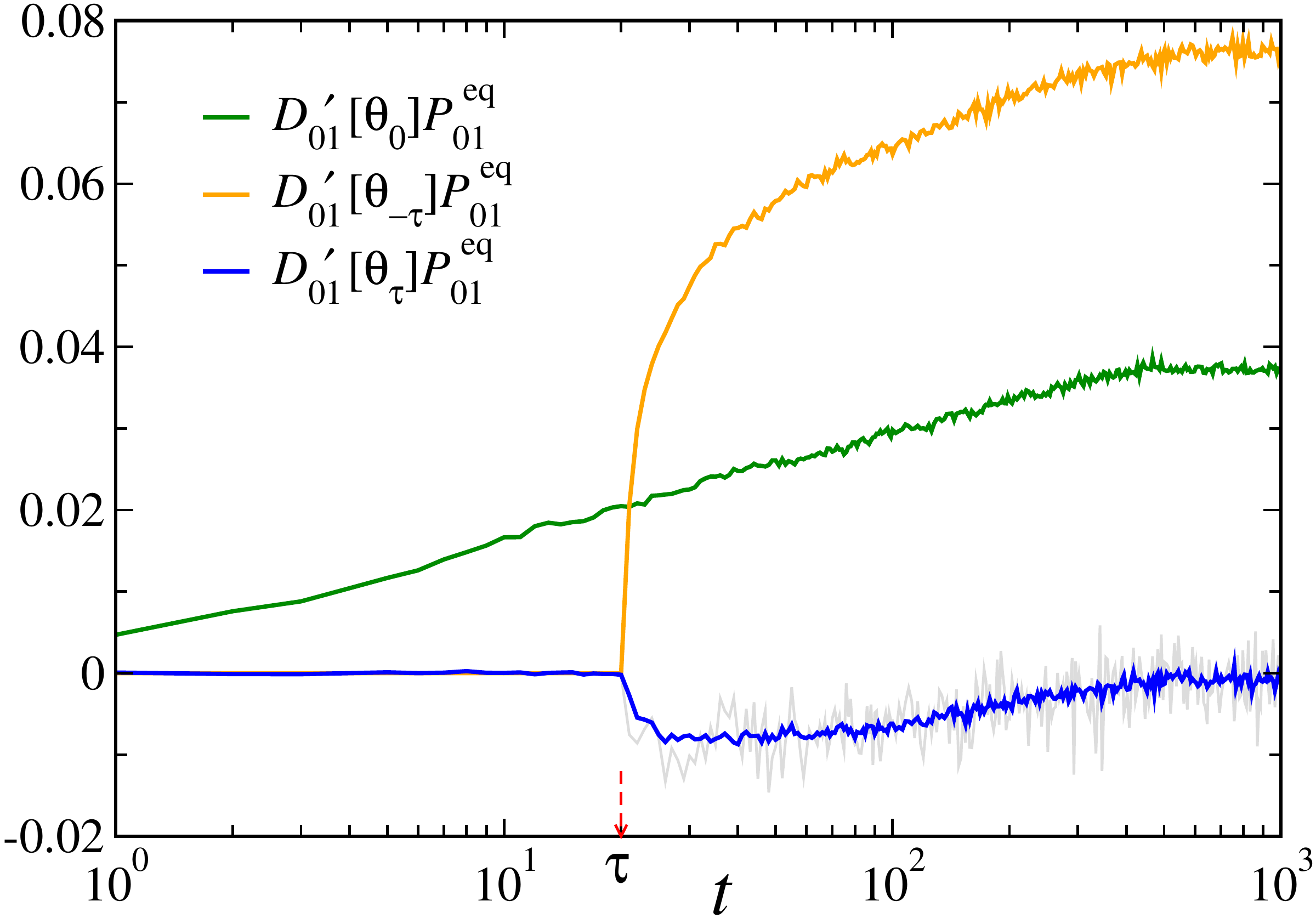}
	\caption{The relevant time-symmetric components $D_{01}^\prime$ contributing to the second order response for the Ising model measured from numerical simulations. Here we have considered a fixed $\tau=20,$ and the linear response is calculated using $\varepsilon=0.05.$ The blue curve is obtained using the best fit for $P_{10}^\varepsilon[\theta_{t-\tau}]$, while the grey curve shows the original data (see the main text and Appendix \ref{app:ising} for details).}\label{fig:ising_Dprime}
\end{figure}

For the sake of simplicity, we take  the two-step protocol introduced in Sec.~\ref{sec:twostep} [see Eq.~\eqref{eq:h_2step}] with $\Delta h_0=\Delta h_1=1$ and consider the response of the observable $O(X)=X.$  We also use a time independent offset magnetic field of strength $g=2.0$, which renders the equilibrium system non-symmetric, yielding a finite second order response.

  Unlike the four-state model, the susceptibility and response function cannot be calculated  analytically here and we take recourse to Monte-Carlo simulations. To be specific, we use Glauber dynamics, where a randomly selected spin flips with rate $\min \{1, e^{-\beta \Delta H}\},$ $\Delta H$ being the change in energy due to the proposed flip and $\beta=T^{-1}$ is the inverse temperature of the system. One Monte-Carlo step consists of $L^2$ attempted flips, which defines the unit of time.

 To demonstrate the validity of the response formalism, we compare the  response  $\la O \ra^{(2)}(t)$ predicted by Eq.~\eqref{eqn:assumption} with $\la O \ra^{(2)}_\text{per}(t),$ obtained from directly applying a larger perturbation. The latter is extracted accurately from,
  \bea 
 \la O \ra ^{(2)}_\text{per}(t) = \frac1{2 \varepsilon^2}[\la X\ra^{\varepsilon}(t) + \la X\ra^{-\varepsilon}(t) - 2 \la X \ra^\text{eq}],
 \eea  
 where $\la \cdot \ra^{\varepsilon}$ denotes the expectation value in the presence of the perturbation protocol of Eq.~\eqref{eq:h_2step} with strength $\varepsilon$ and $\la X \ra^\text{eq}$ is the expected value in equilibrium. We use measurements with strengths $\pm \varepsilon$ to avoid errors of $O(\varepsilon^3)$ \cite{basuextrapolation}.  
 
 On the other hand, the response theory  predicts the second order susceptibility via Eq.~\eqref{eqn:assumption}. For the protocol of Eq.~\eqref{eq:h_2step} with $\Delta h_0=\Delta h_1=1$ it reduces to,
 \bea 
 \la O \ra ^{(2)}(t) =  \chi(t,t) + 2  \chi(t,t-\tau) +  \chi(t-\tau,t-\tau)\; \label{eq:O2_predicted}
 \eea 
 where $\chi(t_1,t_2)$ is given by Eq.~\eqref{eqn:twoindex}. As mentioned before, for $O(X)=X,$ the sum reduces to only one term, namely, $i=0,j=1.$ Moreover, in this case, $S^\prime_{01}= \beta(V(1)- V(0))=2 \beta,$ and we only need to measure the linear parts of $D_{ij}$ under single-step perturbations at times $0$ and $\pm \tau.$

Using the Monte-Carlo simulations and applying a (small) perturbation of strength $\varepsilon=\pm 0.05,$ we measure the {\it linear responses} of the relevant path probabilities $P_{ij}[h](t)$.
The corresponding matrices $D'$ are then computed using Eq.~\eqref{eq:D_epspm} in the Appendix \ref{app:ising}. Figure~\ref{fig:ising_Dprime} shows plots of the resulting  $D_{01}^\prime$, evaluated for the three different protocols as needed, in analogy to Fig.~\ref{fig:D01}.  Qualitative differences to  Fig.~\ref{fig:D01} result from the fact that here, a finite second order response remains in the long-time limit. The presence of a slow time-scale is visible in the slow relaxation of the curves in Fig.~\ref{fig:ising_Dprime}.

For the particular quantity $P_{ji}[\theta_{t-\tau}]$ the numerical fluctuations are substantial and we therefore obtain the derivative by fitting the  $P_{ji}^{\pm \varepsilon}[\theta_{t-\tau}]$ to a compressed exponential form and taking the difference of these fitted functions; see Appendix \ref{app:ising} for more details. The dark blue curve shows the $D^\prime_{01}[\theta_\tau]$ obtained using this fit; the light grey curve shows the original data.

\begin{figure}[t]
	\centering
	\includegraphics[width=7.5 cm]{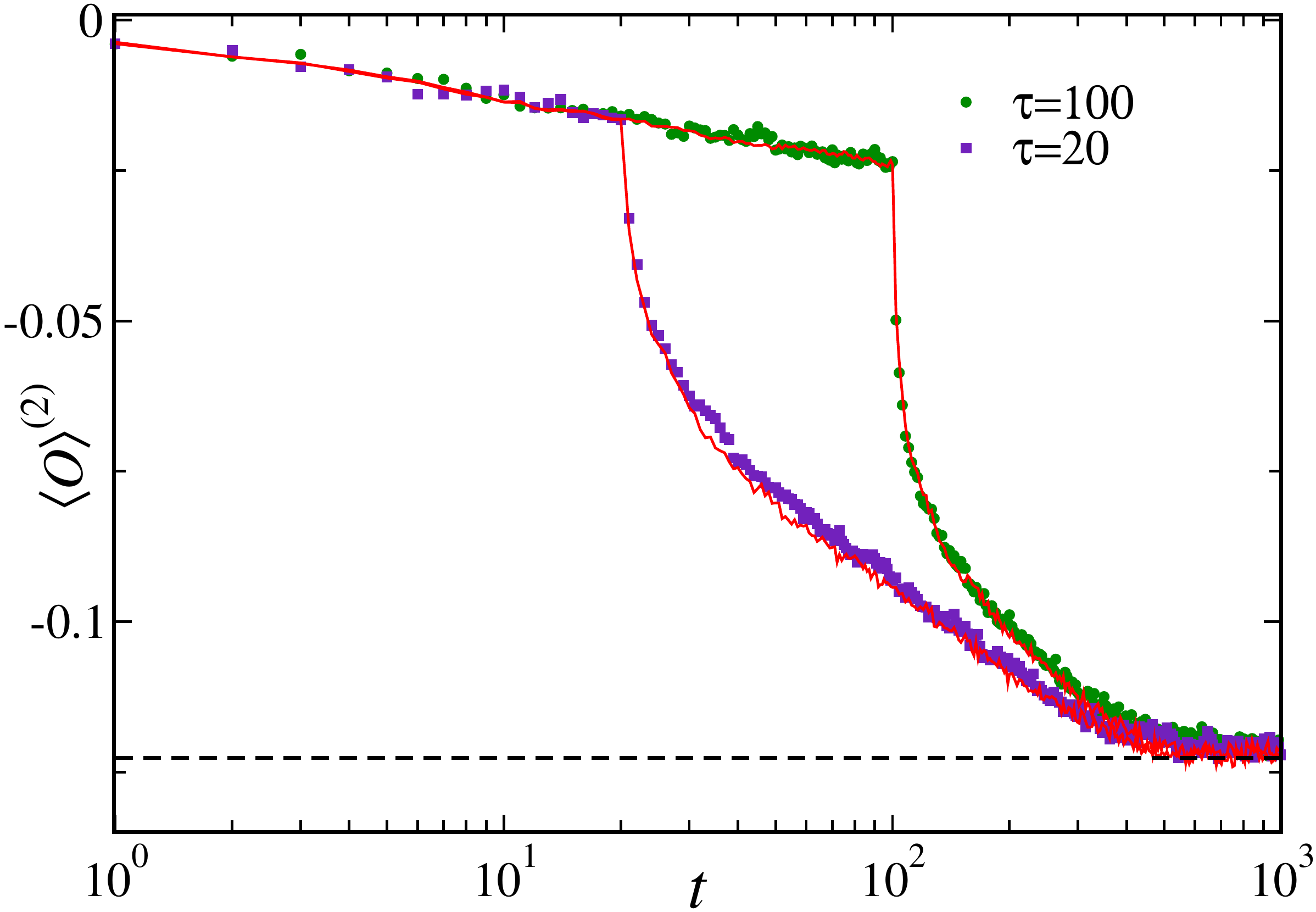}
	\caption{The second order response of a single spin in the Ising model. The solid red lines show the response $\la O \ra ^{(2)}$ predicted from Eq.~\eqref{eq:O2_predicted} while the symbols show the susceptibility $\la O\ra^{(2)}_\text{per}$ measured directly from simulations. The dashed line indicates the {\it static} response, found by expanding the Boltzmann weight, see main text.}\label{fig:ising_chi}
\end{figure}

The second order response is obtained using Eq.~\eqref{eq:O2_predicted} along with 
 Eq.~\eqref{eqn:twoindex}.  Figure~\ref{fig:ising_chi} compares the susceptibility $\la O \ra^{(2)}_\text{per}(t)$ that we measure directly (symbols) with the predicted response $\la O \ra^{(2)}(t)$ (solid lines) for two different values of $\tau.$  At late times $t \to \infty,$ the susceptibility reaches a stationary value that is independent of $\tau$ and is nothing but the equilibrium second-order response for a perturbation $\varepsilon (\Delta h_0 + \Delta h_1) V(X) = 2 \varepsilon (2 X-1).$ This can be calculated by a series expansion of the Boltzmann weight and turns out to be $8 \beta^2 \la X \ra^\text{eq}(1-2\la X \ra^\text{eq})(1-\la X \ra^\text{eq})$  as shown in detail in Appendix \ref{sec:static}. This value is indicated by a black dashed line in the figure. 
  
 It is worth re-emphasizing here that the procedure used in this section generalizes  
 the extrapolation scheme introduced in Ref.~\cite{basuextrapolation} to arbitrary time-dependent perturbations: the second order response, which is relevant for a comparatively stronger perturbation, can be predicted exactly by measuring path probabilities close to equilibrium, i.e.~within the linear response regime.

\section{Conclusions}
We have developed a second order response theory for coarse grained observables, which is valid for arbitrary time dependent perturbation protocols and thus provides a significant extension of Ref.~\cite{basuextrapolation}. One application of this theory is an extrapolation scheme that uses measurements within the linear regime to predict the second order response. The relevant linear measurements only need to be performed for simple perturbation protocols consisting of a single switch-on event, and from these the second order response for arbitrary protocols follows. 

The necessary spatial resolution, i.e.~the degree of coarse graining possible in this approach, is set by the perturbation. Returning to the introductory example of two colloidal particles, if the perturbation acts only on one of the two colloids, the other one can be coarse grained, i.e.~its position does not have to be monitored. An important difference to approaches which use fast and slow variables is thus that, in the present scheme, the coarse grained variables are allowed to be of non-Markovian type. 

The  scheme can be applied to any {\it time} dependence of protocol. As is the case for spatial resolution, it is the protocol that sets the (experimental) time resolution required to apply our method. However, as found in explicit examples, a temporal resolution that is fine compared to the reaction time of the coarse grained variables is also sufficient.  

Technically, our approach relies on being able to resolve fully the  entropy production. This is ensured by the assumption that the perturbing potential depends only on macrovariables, and it implies that the entropy production and the time symmetric part of the action decouple when coarse graining.
This work is thus naturally in agreement with (macroscopic and stochastic) thermodynamics and with the known fluctuation relations. Its new contribution lies in the description of the non-thermodynamic, symmetric part of the action.

Future work will consider higher orders of perturbation, as well as simultaneous perturbations by multiple potentials with different time dependences, and will investigate possibilities of combining this scheme with approaches that rely on separation of fast and slow time scales. 
It may also be insightful to combine our approach with estimates of the entropy production for cases where the potential acts on partly inaccessible d.o.f.~\cite{alemany2015free}.
Another important extension to be addressed is perturbations of non-equilibrium systems, where already the linear order is difficult as regards coarse graining.

Finally, one could explore the question of a  Gallavotti-Cohen symmetry for coarse grained descriptions~\cite{crooks1999,Lebowitz99}: from the definition of $s_{\varepsilon,h}(\omega)$ and Eq.~\eqref{eqn:macroent} we see that $\langle e^{-s_{\varepsilon,h}(\omega)}\rangle = \langle e^{-\varepsilon S_h'(\Omega)}\rangle = 1$. Because the entropy production $\varepsilon S_h'(\Omega)$ depends only on the coarse-grained paths, this implies that in our setting the coarse-grained dynamics does itself obey the Gallavotti-Cohen symmetry. It will then be interesting to see whether statements about nonlinear response can be deduced from this, by extending existing results for the linear response regime~\cite{crooks1999,Lebowitz99}. 
\acknowledgments

 U.~B.~acknowledges support from Science and Engineering Research Board (SERB), India under Ramanujan Fellow-ship (Grant No.  SB/S2/RJN-077/2018). M.K. acknowledges support from DFG Grant No. KR 3844/3-1.

\appendix
\section{Relations and definitions}\label{sec:ApA}
\subsection{Microscopic response formalism}
The results presented here are based on the known microscopic response formalism presented in section \ref{theory}. In this appendix, we aim to describe in detail how time dependent perturbations can be treated, as these details may not be provided in other literature. 
The key is the reversal of paths and protocol in the integration.
Mathematically, the path weight considered here is a joint one.
This is in contrast to some common literature on the subject of path integrals, which considers the probability $p(\omega|x_i)$ to find a path $\omega$ given that the system starts in a fixed state $x_{i}$. This probability is related to the full path probability by the probability $\rho_0(x_i)$ of the initial state at time $s=0,$  namely $p(\omega) = p(\omega|x_i) \rho_0(x_i)$.
In fact, one may think of this path integral as three integrals: one Lebesgue integral over the starting point $x_0$, one over the end point $x_t$ and a third real path integral over possible paths connecting these two points.
It is then immediately clear what happens when integrating over reversed paths in the average of a state observable as given by Eq.~\eqref{eqn:av}:
We can integrate out paths starting in a fixed state $x_0 = x_i$ at which the observable $O(x_0)$ is then evaluated.
Since every path starting in $x_i$ has to go somewhere, the integral over all these paths reduces to the probability density $\rho_0(x_i)$ of $x_i$:
\begin{align}
        \int_{x_0 = x_i} p_{\varepsilon,g}(\omega) \mathbb{D} \omega = \rho_0(x_i)~.
\label{app:rel}
\end{align}
Note that the same integral, taken over the {\em conditional} path weight, would yield one.
These arguments hold for \emph{arbitrary protocols}, and in particular for the reversed protocol $\overline{h}(s) = h(t-s)$.

\subsection{Coarse-grained path integral\label{app:pi}}
Let us now consider the \emph{coarse-grained path integral} as used in Eq.~(\ref{eqn:chi2coarse}) in more detail.
Instead of expressing the coarse-grained path integral by  summing 
over possible states, see Eq.~\eqref{eqn:cgint}, it can also be written by integration over possible paths as follows. 
 
For processes with discrete state spae, paths are given by a sequence of states and jump times $\Omega =(X_0, t_0; X_1, t_1, \dots, X_n, t_n)$,
i.e.~$X_s = X_k$ for all $s \in [t_k, t_{k+1})$. We always have $t_0 = 0$ and we set $t_{n+1}=t$, for notational simplicity.
A path integral is then given by summing over possible states and integrating over jump times $t_i$.
\begin{align}
\begin{split}
	&\int f(\Omega) \mathbb{D} \Omega = \sum_{n =0}^{\infty}
	\sum_{X_0 \in \Gamma'}\sum_{X_1 \in \Gamma'\setminus \{X_0\}} \ldots \sum_{X_{n} \in \Gamma'\setminus \{X_{n-1}\}}\\
	&\int_0^{t} \int_{t_1}^t \ldots \int_{t_n}^t ~ f(\Omega)~   \prod_{i =0}^{n-1}  \d t_{i+1} ~.
      \end{split}
    \end{align}
Analogously one may write down the exact same equation for Markov jump processes, as done in more detail in Ref.~\cite{wynantsthesis}.

\subsection{Derivation of Eq.~\eqref{eqn:twoindex}}
For clarity, we provide a few more steps for the derivation of the exact second order susceptibility in terms of sums, Eqs.~\eqref{eqn:1} and \eqref{eqn:twoindex}.
We insert the linear decomposition Eq.~\eqref{eqn:D} of the dynamical activity $D_{ikj}'$ into the second order response $\av{O}^{(2)}$ expressed as a sum over states, Eq.~\eqref{eqn:twosteps}.
For a protocol with two steps, as defined in section \ref{sec:twostep}, separating the different combinations of $\Delta h_0$ and $\Delta h_1$ yields
\begin{align}
  \begin{split}
    & \av{O}^{(2)}= \\
    &- \sum_{ikj} \big(\Delta h_0^2 S_{ij}' D_{ikj}'[\Theta_0] 
   + \Delta h_1^2 S_{kj}' D_{ikj}'[\Theta_{\tau}]\\  
     & +\Delta h_0 \Delta h_1  S_{kj}' D_{ikj}'[\Theta_0]
    + \Delta h_0 \Delta h_1 S_{ij}'  D_{ikj}'[\Theta_{\tau}]
  \big)P_{ikj}^{\rm eq} O(j)~.
\end{split}
\label{eqn:app2}
\end{align}
Adding the explicit time arguments, the two versions of $D'$ appearing in Eq.~\eqref{eqn:D} and Eq.~\eqref{eqn:app2} read
\begin{align}
  \notag &D_{ikj}'[\Theta_0](\tau,t) P_{ikj}^{\text{eq}}(\tau,t)=-\frac{1}{2}
 \biggl(P_{ikj}'[\Theta_0](\tau,t)\\
       &+ P_{jki}'[\Theta_0](t-\tau,t)\biggr)\label{eqn:D0}\quad \text{and}\\
\notag &D_{ikj}'[\Theta_{\tau}](\tau,t) P_{ikj}^{\text{eq}}(\tau,t)=-\frac{1}{2}
       \biggl(P_{ikj}'[\Theta_{\tau}](\tau,t)\\
       &+ P_{jki}'[\Theta_0](t-\tau,t)
     -P_{jki}'[\Theta_{t-\tau}](t-\tau,t)\biggr)\label{eqn:Dt}
\end{align}
In Eq.~\eqref{eqn:Dt} we have expanded the probability under the backwards protocol $\overline{\Theta_\tau} = \Theta_0 -\Theta_{t-\tau}$ by thinking of it as a two-step perturbation and using the general linearity property
\begin{align}
      P_{ikj}'[h^{(2)}]&=\Delta h_0 P_{ikj}'[\Theta_0](\tau,t) + \Delta h_1 P_{ikj}'[\Theta_{\tau}](\tau,t)~.
\end{align}
The above equations are the basis for integrating out one index, specifically the  initial state $i$ in Eq.~\eqref{eqn:D0} and the state $k$ at time $\tau$ in Eq.~ \eqref{eqn:Dt}.
Summing over possible states in a joint probability yields $\sum_j \mathcal{P}(X_{t_1} = i,X_{t_2} = j, X_{t_3} = k) = \mathcal{P}(X_{t_1} =i, X_{t_3} = k)$, thus yielding probabilities $P_{ij}$ for different protocols, see Eq. \eqref{eqn:sumup0} and \eqref{eqn:sumup1} respectively.

\subsection{Markov case \label{sec:markov}}

The results derived in sections \ref{sec:coarse} and \ref{sec:conc} do not rely on a Markovian property of the coarse grained variables. 
There might be practical cases, however, where the degrees of freedom under consideration {\em are} in fact Markovian, for example if a local equilibrium approximation is justified for the d.o.f.~that are integrated out.
In that case, $(X_s)_{s\in[0,t]}$ is a Markov process, and hence follows the formulas of the microscopic response formalism in section \ref{sec:micro} (see Refs.~\cite{colangeli2011meaningful,basufrenetic} for specifics). Notably, the linear contribution of the time symmetric part is given as a superposition of instantaneous values, denoted $\tilde{d}(x_s)$,
\begin{align}
  d_h'(\omega) = \int_0^t h(s) \tilde{d}(x_s) \d s
\end{align}
as explained in Ref. \cite{colangeli2011meaningful}. We can thus decompose
\begin{align}
 & D_{ikj}'[\Delta h_0 \Theta_0 + \Delta h_1 \Theta_{\tau}] = \Delta h_0 D_{ik}'(\tau) +\Delta h_1 D_{kj}'(t-\tau)
 \end{align}\label{eqn:DM}
 and for the probability
 \begin{align}\label{eqn:CK} 
 P_{ikj} = P_{ik}(\tau) p_{kj}(\tau,t)
\end{align}
with the conditional probablity $p_{kj}$ introduced before Eq.~\eqref{eqn:master}.
Eq.~\eqref{eqn:CK} is in contrast to the case of non-Markovian processes where states at different times couple due to memory effects.
With only the quantity $D_{ik}'[\Theta_0]$  appearing in Eq.~\eqref{eqn:DM} (evaluated at different times), the second order response in the Markov case thus takes the complexity of the single step protocol described in Sec.~\ref{sec:si}. This simplifies the extrapolation scheme introduced in Section \ref{sec:recipe}, as only $P_{ij}'[\Theta_0]$ and  $P_{ij}^{\rm eq}$ need to be measured in order to find the second order response for any protocol.

\section{Models and measurement}
\subsection{The four state model}\label{app:4st}
The rate matrix for the four state model analyzed in section \ref{sec:fourstate} is given by
\begin{align}
  q(s) = \begin{pmatrix}
 -r & r & 0 & 0\\
 r & -e^{\varepsilon h(s)}-r & e^{\varepsilon h(s)} & 0\\
0 & 1 & -1-r & r &0\\
0 & 0 & r & -r
  \end{pmatrix}
  \label{eqn:rates}
\end{align}
so that the row sums are 0 and as explained $r$ is a dimensionless parameter.
For $r \ll 1$ this system exhibits much slower rates within the macrostates than in the transition connecting the two coarse-grained states, which is of order 1. 
Still our extrapolation technique succeeds (Fig. \ref{fig:3steps}).
This illustrates that our method does not rely on separation of timescales as also demonstrated in Ref.~\cite{basuextrapolation}.
For the average of the coarse-grained observable $O(X) = X$ the second order susceptibility is computed from Eq. \eqref{eqn:k}.

Using the abbreviations $\gamma = \sqrt{1+r^2}$ and $\gamma_{\pm} = 1+r\pm \gamma$
the second order susceptibility reads
\begin{align}
  \begin{split}
	 \chi(t_1,t_2)&= 
	 \frac{1}{64 \gamma^3} \bigg( r\left( e^{2 t_2\gamma} -1 \right) \\
	   &\big(e^{-t_1\gamma_+}+2e^{-t_2 \gamma_+} + e^{-2 t_2\gamma-t_1\gamma_-} \big)\\
	&+2t_2\gamma \big(\left(1+\gamma \right) (\gamma-r)\left( e^{-t_1\gamma_+}+e^{-t_2\gamma_+} \right) \\
	&+\left( \gamma-1 \right)\left( \gamma+r \right) \left(e^{-t_2\gamma_-}+ e^{-t_1\gamma_-}  \right)  \big)
\bigg)~.
\end{split}
\label{eqn:kfour}
\end{align}

\subsection{Measurement in the Ising model}\label{app:ising}

To compute the first derivatives of the path weight $P_{ij}[h](t)$ accurately,
we use two measurements, namely, with perturbation strengths $\pm \varepsilon.$ Expanding $P_{ij}^{\pm \varepsilon}[h](t)$ in a Taylor series around equilibrium, i.e.~around $\varepsilon=0$, 
we get
\bea 
P_{ij}^{\pm \varepsilon}[h](t) =  P_{ij}^\text{eq}(t) \pm  \varepsilon P_{ij}'[h](t) + \frac{\varepsilon^2}{2} P_{ij}''[h](t) + {\cal O}(\varepsilon^3).\nonumber
\eea 
The first derivative of the path probabilities can then be extracted from,
\bea 
P_{ij}'[h](t) = \frac 1{2\varepsilon} \left[ P_{ij}^{\varepsilon}[h](t) - P_{ij}^{-\varepsilon}[h](t) \right] + {\cal O}(\varepsilon^2), \label{eq:Pij_prime}
\eea 
and similarly for $P_{ij}[\bar h](t).$ Note that the error here is one order smaller than if the derivative was computed only from $P_{ij}^{\varepsilon}[h](t)$ and $P_{ij}^\text{\rm eq}(t).$ Using the above equation and its counterpart for $P_{ij}[\bar h](t)$ we can also extract $D_{ij}',$  
\bea 
D_{ij}'[h]P_{ij}^{\text{eq}} &=& -\frac{1}{4 \varepsilon} \left( P_{ij}^\varepsilon[h] +P_{ji}^\varepsilon[\overline{h}]- P_{ij}^{-\varepsilon}[h] -P_{ji}^{-\varepsilon}[\overline{h}]\right) \nonumber \\[0.25 em]
&& + \mathcal{O}(\varepsilon^2). \label{eq:D_epspm}
\eea 

As mentioned in the main text, for the particular case of  $P_{ij}[\theta_{t-\tau}](t),$ instead of calculating the derivative directly from the numerically measured path probabilities,  we use a functional fit. 
We first fit $P_{10}^{\varepsilon}[\theta_{t-\tau}](t)- P_{10}^{\varepsilon}[\theta_{t-\tau}](\tau) $ to a functional form $ a(1- \exp{[-b (t - \tau)^c]})$ (remember that the path probability difference is zero for $t<\tau$ in this case) with $a,b,c$ as fitting parameters. The derivative is then calculated using Eq.~\eqref{eq:Pij_prime} along with these fitted functions. For the sake of completeness, we provide the values of the fitting parameters in Table \ref{tab:param}. 

\begin{table}[]
    \centering
      \begin{tabular}{c|c|c|c|c}
      \hline
     & $\varepsilon$  &  $a$ & $b$ & $c$ \\
     \hline 
   \multirow{2}{*}{$\tau=20$}& 0.05 & 0.0238 & 0.0296 & 0.7752 \\
    & \; -0.05 \;  & \; 0.0253 \; &\; 0.0327 \;  & \; 0.7575 \; \\
    \hline 
   \multirow{2}{*}{$\tau=100$}& 0.05  & 0.0096 & 0.0068 & 1.0227 \\
    &  -0.05  & 0.0095 & 0.0045 & 1.0968 \\
    \hline 
\end{tabular}
 \caption{Numerical values of the fitting parameters used for $P_{10}^{\pm \varepsilon}[\theta_{t-\tau}](t)$}
    \label{tab:param}
\end{table}

\subsection{Static response in the Ising model \label{sec:static}}

The long-time limiting value of the second order response in  the Ising model can be computed from the equilibrium Boltzmann distribution. Under the perturbation protocol \eqref{eq:h_2step}, in the long-time limit, the system reaches an equilibrium state characterized by configuration weights
\bea
P(\{\eta_i\}) = \frac 1{Z_\varepsilon} e^{-\beta [{\cal H}_0-\varepsilon(\Delta h_0 + \Delta h_1) V(X)]}
\eea 
where $Z_\varepsilon$ is the equilibrium partition funcion and  ${\cal H}_0$ is the Hamiltonian in the absence of the perturbation.  The second order response of any observable $O$ can be calculated by expanding the above weight around $\varepsilon=0,$ multiplying by $O$ and summing over all possible configurations. This straightforward excercise leads to a formal expression,
\bea
\la O \ra^{(2)} &=& \frac {\beta^2}2 (\Delta h_0 + \Delta h_1)^2  \Big [\la O V^2\ra - \la O\ra \la V^2 \ra \cr
&& + 2 \la O \ra \la V\ra^2-2 \la O V \ra \la V\ra \Big]\nonumber 
\eea 
For the case $O(X)=X$ and $ V(X)= 2 X -1$ with $\Delta h_0=\Delta h_1=1$  the above expression simplies to
\bea 
\la O \ra^{(2)} = 8 \beta^2 \la X\ra^{\text{eq}}(1- 2 \la X\ra^{\text{eq}})(1-\la X\ra^{\text{eq}})
\eea 
where we have used the fact that $X^2=X.$

\bibliography{libforintro.bib}

\end{document}